\documentclass[aps,prl,twocolumn,showpacs,amsmath,amssymb,superscriptaddress]{revtex4}
 
\usepackage{graphicx}% Include figure files,superscriptaddress,
\usepackage{bm}% bold math
\usepackage{amssymb}
\usepackage{hyperref}

\def\beq{\begin{eqnarray}}
\def\eeq{\end{eqnarray}}
\def\bec{\begin{center}}
\def\enc{\end{center}}
\def\bit{\begin{itemize}}
\def\eit{\end{itemize}}

\begin{document}

\title{Generating far-field orbital angular momenta from near-field optical chirality}

\author{Yuri Gorodetski}
\affiliation{ISIS, Universit\'{e} de Strasbourg and CNRS (UMR 7006), 8 all\'{e}e Gaspard Monge, 67000 Strasbourg, France}
\author{Aur\'elien Drezet}
\affiliation{Institut N\'eel, UPR 2940, CNRS-Universit\'e Joseph Fourier, 25, rue des Martyrs, 38000 Grenoble, France}
\author{Cyriaque Genet}
\affiliation{ISIS, Universit\'{e} de Strasbourg and CNRS (UMR 7006), 8 all\'{e}e Gaspard Monge, 67000 Strasbourg, France}
\author{Thomas W. Ebbesen}
\affiliation{ISIS, Universit\'{e} de Strasbourg and CNRS (UMR 7006), 8 all\'{e}e Gaspard Monge, 67000 Strasbourg, France}

\begin{abstract}
We demonstrate that nanostructures carefully designed on both sides of a thin suspended metallic membrane
couple light into a chiral near field and transmit vortex beams through a central aperture that connects the two sides of the membrane. We show how far-field orbital angular momentum (OAM) indices can be tailored through nanostructure designs. We reveal the crucial importance of OAM selection rules imposed by the central aperture and derive OAM summation rules in perfect agreement with experimental data. 
\end{abstract}

\pacs{}

\maketitle

%%%%%%%%%%%%%%%%%%%%%%%%%%%%%%%%%%%%%%%%%%%%%%%%%%%%%%%%%%%
Structured light beams with phase or polarization singularities, have revealed unique optical 
properties with applications ranging from super-resolution imaging \cite{HellNatMeth2009} to 
high-resolution sensing \cite{DemasOptLett2012} and from particle micromanipulation \cite{Dunlop} to 
quantum optics \cite{Torres}. Currently, chiral nanostructures draw promising 
routes for enhancing singular optical signatures, thus providing extended control over new 
functionalities in metamaterial science \cite{Pendry} and biomimetics engineering \cite{ParkerNatNanotech2007,GuOptX2011}. Interestingly, while the connection between optics and chirality is 
well established for 3 dimensional (3D) chiral structures, the interaction of chiral light with 2D chiral objects is a 
topic of on-going debate \cite{Zheludev,BaiPRA2007,Drezet}, with strong potential in physical chemistry for chirality enhancement in the near field \cite{Cohen,HendryNatNano2010,GiessenNanoLett2012}.
 
Recently, singular optical effects have been discussed in the near field, in particular 
in relation to chiral surface plasmon (SP) modes which have been shown to carry orbital 
angular momentum (OAM) \cite{Babiker,Zhan2009,Halas,Garcia-Vidal}. But until now, singular SP modes and associated spin-orbit coupling have only been probed in the near field \cite{ChoOptX2012,Ohno,PRL2008}. Studies on plasmonic beaming with OAM have been scarce \cite{Nano2009,Zhan} and 
the relation between near-field chirality and OAM in the far field was never addressed.

In this Letter, we demonstrate that nanostructures carefully designed on both sides of 
a thin suspended membrane lead to tailoring optical OAM in the far field. Single and double-
sided plasmonic structures consisting of concentric grooves periodically spaced from a 
central subwavelength aperture -- so called plasmonic bull's eye (BE)- have shown extraordinary 
optical transmission and beaming effects \cite{Beaming_Science,DegironOptX2004}. Yet, the OAM
behaviour of these structures was not discussed. This Letter presents a comprehensive analysis of 
the OAM transfer during plasmonic in-coupling and out-coupling by chiral nanostructures at each 
side of a membrane, stressing in particular the role of a back-side structure in generating vortex
beams as $e^{i\ell\varphi}$ with tunable OAM indices $\ell$. With such membranes, we achieve propagating beams carrying 
OAM up to $|\ell | =  8$. Moreover we reveal and analyze the fundamental role of the
central aperture in the system's OAM generation through specific selection rules.  

Our device consists of a suspended thin ($h\sim 300$ nm) metallic membrane, fabricated by evaporating 
a metal film over a poly(vinyl formal) resine supported by a transmission electron microscopy 
copper grid. After evaporation, the resine is removed using a focused ion-beam (FIB), leaving a 
freely suspended gold membrane. Plasmonic structures are milled, in either concentric (BE) or spiral
geometry on both sides of the membrane around a unique central cylindrical aperture acting as 
the sole transmissive element of the whole device. 
%%%%%%%%%%%%%%%%%%%%%%%%%% FIGURE 1 %%%%%%%%%%%%%%%%%%
 \begin{figure}[ht]
 %\vspace{1cm}
\includegraphics[width=7.5cm, keepaspectratio] {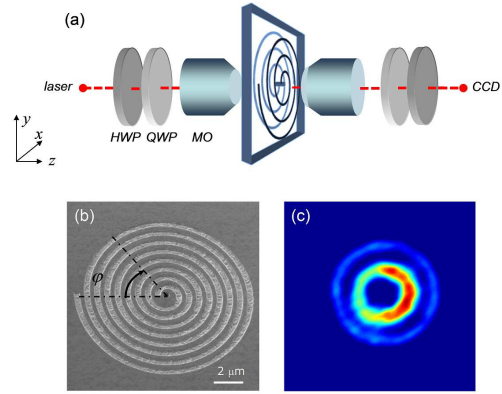}%
 \caption{(a) Experimental setup: the incoming laser beam is circularly polarized using half (HWP) and quarter-wave (QWP) plates and weakle focused by a microscope objective ($5\times$, NA$=0.13$). The transmitted beam is imaged by a second objective ($40\times$, NA=$0.60$) and a lens tube ($f=200$ mm, not shown) on a CCD camera and analyzed in the circular polarization basis by additional HWP and QWP. (b) Scanning electron microscope image
of a $L_1$ spiral milled on the Au membrane, with $m=-1$ and $\lambda_{\rm SP}=768$ nm. (c) Intensity distribution imaged through a BE-$L_1$ structure illuminated with an right circular polarization and analyzed with a left circular polarization.}
 \label{fig1}
 \end{figure}
%%%%%%%%%%%%%%%%%%%%%%%%%%%%%%%%%%%%%%%%%%%%%%%%%%%%%%
The general groove radial path is given in the polar $(\hat{\boldsymbol \rho},\hat{\boldsymbol \varphi})$ basis, 
as ${\boldsymbol \rho}_{n}=(n\lambda _{\rm SP}+m\varphi \lambda _{\rm SP}/2\pi)\hat{\boldsymbol \rho}$, with 
$n$ an integer, $\lambda _{\rm SP}$ the SP wavelength and $m$ a pitch number. We start with the simple 
situation of a plasmonic BE structure ($m = 0$) milled on the front-side of a gold membrane and a spiral with 
a pitch of $m=1$ on the back-side. Note that orientation conventions are chosen with respect to the light 
propagation direction, so that a right-handed spiral $R_{m}$ corresponds to $m > 0$ and a left-handed spiral 
$L_{m}$ to $m < 0$. 

As shown in Fig.\ref{fig1}, the membrane is illuminated by a single mode fiber pigtailed laser diode at $\lambda_0=785$ nm. 
The depth of the grooves ($30$ nm) is smaller than the skin depth ($\sim70$ nm at $\lambda_0$) so that light on the 
front-side is only transmitted through the central hole perforating the membrane. We start with an aperture of subwavelength 
diameter $400$ nm that still provides sufficient signal-to-noise ratio when imaging in the far field the transmitted beam. 
We take a paraxial incoming beam ${\bf E}^{\rm in}_\pm({\boldsymbol \rho},z)e^{-i\omega t}=\hat{\boldsymbol \sigma}_\pm{\cal E}({\boldsymbol \rho})e^{ik_z z}e^{-i\omega t}$ with an incident wavevector $k_z = 2 \pi/\lambda_0$, in either right ($\hat{\boldsymbol \sigma}_+$) or left ($\hat{\boldsymbol \sigma}_-$) circular 
polarization state. The transmitted beam is analyzed in the same circular basis and far-field images ${\cal M}_{ij}$ are 
recorded, with the four $(i,j)=(\pm,\pm)$ combinations of preparation $j$ and analysis $i$.

\begin{figure}[ht]
 %\vspace{1cm}
 \includegraphics[width=7.5cm, keepaspectratio] {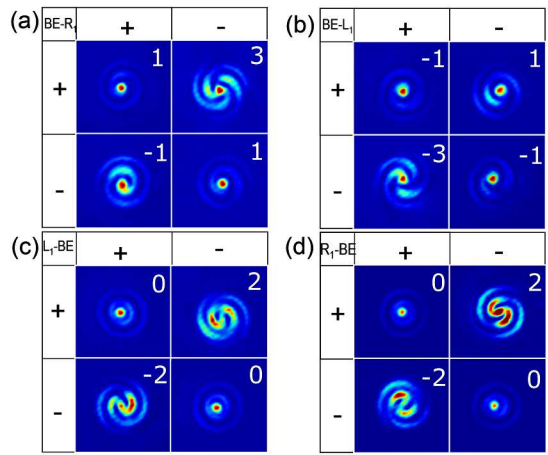}%
 \caption{Intensity distributions of the beam emerging from (a) BE-$L_1$, (b) 
BE-$R_1$, (c) $R_1$-BE and (d) $L_1$-BE structures, respectively. 
The hole diameter used in all the structures was $400$ nm. 
Labels ($\pm, \pm$) correspond to the combination of circular polarization preparation and analysis. 
The numbers correspond to the corresponding OAM indices.}
 \label{fig2}
 \end{figure}

Fig.\ref{fig2} displays such transmission images for membranes comprising a BE in the front-side and 
alternatively $L_1$ and $R_1$ spirals on the back-side. The spiralling structure of the images can be 
understood as the Fano-like interference between the spherical wavefront of the light 
directly diffracted through the hole and the quasi-planar wavefront of the vortex beam decoupled from the plasmonic nanostructure on the back-side of the membrane. 
Our system has therefore a built-in reference wave that generate the recorded interferograms whose 
number of arms and handedness directly give the OAM index and sign $\pm\ell$ of the vortex beam \cite{Fringes,note1,Allen_book}. 
This way, we measure OAM values of $\ell_{++} = \ell_{--} = -1$ and $\ell_{-+} = 1$ and $\ell_{+-} = -3$ for the 
crossed-terms shown in Fig.\ref{fig2} for the BE-$R_1$ configuration. The OAM signs are merely reversed 
when a $L_1$ spiral is milled on the back-side, as shown in Fig.\ref{fig2} b).

Near-field generation of OAM at the front-side ($z=0^+$) of the structured membrane can be modeled 
by considering that each point ${\boldsymbol \rho}_n$ of the groove illuminated by the incoming 
field is an SP point source, launching an SP wave perpendicularly to the groove. With groove widths 
much smaller than the illumination wavelength, the in-plane component of the generated SP field in 
the vicinity of the center of the structure is ${\bf E}^{\rm SP}({\boldsymbol \rho}_0,z=0^+)\propto {\sf G}\cdot [{\bf E}^{\rm in}({\boldsymbol \rho}_n,z=0^+)\cdot\hat{\bf n}_n]\hat{\bf n}_n$ where ${\sf G}=e^{ik_{\rm SP}|{\boldsymbol \rho}_0-{\boldsymbol \rho}_n|}/(|{\boldsymbol \rho}_0-{\boldsymbol \rho}_n|)^{1/2}$ is the Huygens-Fresnel plasmonic propagator \cite{LaluetOptX2007,TeperikOptX2009} and $\hat{\bf n}_n=\kappa^{-1} ({\rm d}^2{\boldsymbol \rho}_n / {\rm d}s^2)$ the local unit normal vector determined from the curvature $\kappa$ and the arc length $s$ of the groove.

The resultant SP field is the integral of elementary point sources over the whole groove structure. As indicated by a full evaluation, we can conveniently limit the integration to radial regions $\rho_n\gg\rho_0$ where the grooves become pratically annular. This leads to $\hat{\bf n}_n \sim -\hat{\boldsymbol \rho}$ and therefore to a simple expression of the integrated SP field ${\bf E}^{\rm SP}\propto\sum_n \sqrt{n\lambda_{\rm SP}}\int\limits_0^{2\pi}{\rm d}\varphi e^{im\varphi}e^{-ik_{\rm SP}\rho_0\cos (\varphi-\varphi_0)}[{\bf E}^{\rm in}_\pm \cdot \hat{\boldsymbol \rho}]\hat{\boldsymbol \rho}$. By rewriting the polarization coupling term $[{\bf E}^{\rm in}_\pm \cdot \hat{\boldsymbol \rho}]\hat{\boldsymbol \rho} = [\hat{\boldsymbol \rho}\otimes\hat{\boldsymbol \rho}]\cdot {\bf E}^{\rm in}_\pm$, the $\otimes$ symbol denoting a dyadic product \cite{Manu}, the in-plane radially polarized SP near field is connected to the incoming field as ${\bf E}^{\rm SP}={\sf C}_{\rm in}\cdot {\bf E}^{\rm in}$ by an in-coupling matrix
\beq
{\sf C}_{\rm in}(m)\propto e^{im\varphi_0}\int\limits_0^{2\pi}{\rm d}\varphi e^{im\varphi}e^{-ik_{\rm SP}\rho_0\cos \varphi}
\hat{\boldsymbol \rho}\otimes\hat{\boldsymbol \rho}
\label{matIn}
\eeq 

Crucial is the $\varphi$-dependency of the circularly polarized illumination $\hat{\boldsymbol \sigma}_{\pm}= 
(\hat{\boldsymbol \rho}\pm i\hat{\boldsymbol \varphi})e^{\pm i\varphi}/\sqrt{2}$ that makes the integration (\ref{matIn}) spin-dependent. The excited SP field hence corresponds to a plasmonic vortex carrying OAM of $\ell_{\rm SP}=m\pm 1$, depending on the incident $\hat{\boldsymbol \sigma}_\pm$ polarization, revealing spin-orbit coupling due to a radial plasmonic structure 
\cite{ChoOptX2012,Ohno,PRL2008,Nano2009}. 

In contrast with these recent studies confined to the near field, we have here the new possibility to decouple the singular near field into the far field with an additional structure on the back-side of the suspended membrane connected to the front-size by the central hole. By symmetry (assuming loss-free unitarity) the out-coupling matrix is simply given as the hermitian conjugate of the in-coupling matrix, i.e. 
${\sf C}_{\rm out}={\sf C}_{\rm in}^{\dagger}$, corresponding to a surface field that propagates away 
from the central hole on the back-side. 

The in -- out-coupling sequence corresponds to the product ${\sf T}={\sf C}^{\dagger}(m_{\rm out})
\cdot {\sf C}(m_{\rm in})$ which, in the circular polarization basis, writes explicitly as
\beq
{\sf T}\propto e^{i(m_{\rm out}-m_{\rm in})\varphi}
\begin{bmatrix}
{\sf t}_{++}     & {\sf t}_{+-}e^{2i\varphi}\\ 
{\sf t}_{-+}e^{-2i\varphi} &    {\sf t}_{--}
\end{bmatrix}
\label{matT}
\eeq
with ${\sf t}_{ij}$ radial functions, detailed in the Supplemental Material section. This expression reveals two contributions: a polarization
dependent geometric phase, within the matrix, that stems from the spin-orbit coupling at the annular groove, 
and a factorized dynamic phase that arises due to the spiral twist of the structure \cite{PRL2008}. Note, 
that for $m_{\rm out}= m_{\rm in} = 0$, ${\sf T}$ describes a pure spin--orbit angular momentum transfer, 
conserving the total angular momentum \cite{PRL2008,Marucci,Brasselet,Manu}. We can cast in a table these results 
in the form of summation rules for the OAM generated and transmitted through the membrane.
\begin{table}[htbp]
	\centering
		\begin{tabular}[t] {|c|c|c|}\hline
		                        & $+$ & $-$ \\ \hline
$+$		&  $m_{\rm out}-m_{\rm in}$            &   $m_{\rm out}-m_{\rm in}+2$         \\ \hline
$-$  	&  $m_{\rm out}-m_{\rm in}-2$          &   $m_{\rm out}-m_{\rm in}$            \\ \hline
	\end{tabular}
	\caption{Far field summation rules for OAM generated through the membrane.}
\label{table}
\end{table}
Remarkably, the measurements presented in Fig.\ref{fig2} are in agreement with this Table, demonstrating 
experimentally OAM transfer to the far field from the excitation of a chiral plasmonic 
near field at the back-side of the membrane.

This analysis however does not exhaust the OAM generation process as it can be plainly seen when flipping 
the membrane. The OAM measurements obtained with a BE-$(L,R)_1$ structures and the (flipped) $(R,L)_{1}$-BE ones 
do not coincide, in contradiction with reciprocity operating on ${\sf T}$. The OAM data obtained e.g., for the 
$L_{1}$-BE configuration are shown in Fig.\ref{fig2} (d) and turn out inconsistent with the expectation values 
of Table \ref{table}. Surprisingly since $L_{1}$ means $m_{\rm in}=-1$, the agreement is reached only when fitting $m_{\rm in}\equiv 0$ in the Table. 
%%%%%%%%%%%%%%%%%%%%%%%%%% FIGURE 2 %%%%%%%%%%%%%%%%%%
%%%%%%%%%%%%%%%%%%%%%%%%%%%%%%%%%%%%%%%%%%%%%%%%%%%%%%
This discrepency points to the pivotal role of the aperture in the process of OAM conservation. 
As it is well known, the modes inside a hole of radius $\rho_h$ and symmetry axis along the $z$ 
direction are classify into transverse electric (TE) and transverse magnetic (TM) modes. The 
associated fields ${\bf E}^{\rm wg}({\boldsymbol \rho},z)$ are simply derived from a scalar potential 
$\psi({\boldsymbol \rho},z)$. Given the symmetry of the problem, a multipolar expansion of this 
potential can be given as $\psi({\boldsymbol \rho},z)\propto\sum_{\ell\geq 0}\sum_{n\geq 1} 
\psi_{\ell n}({\boldsymbol \rho}) e^{i\kappa_{\ell n}z}$ with $\psi_{\ell n}({\boldsymbol \rho})
=J_{\ell}\left(k_{\ell n}\rho\right)e^{i\ell\varphi}$ where $J_{\ell}$ is the $\ell$-order Bessel function of the first kind \cite{Nikitin}. Here $\kappa_{\ell n}^2=k_{\ell n}^2-k^2$ 
is the waveguide propagation wavevector and $k=2\pi / \lambda_{0}$ the wavevector of light. The 
eigenvalues $k_{\ell n}$ are determined from boundary conditions (assuming a perfect 
conductor \cite{Catrysse}) as $k_{\ell n}=x_{\ell n}/\rho_h$ where $x_{\ell n}$ is the $n^{\rm th}$ 
roots of $J_{\ell}\left(x\right)=0$ for a TM mode and $\partial_x J_{\ell}\left(x\right)=0$ for a 
TE mode. There is a lowest (cutoff) value $k_{\ell n}^{\rm c}$ of $k$ for which $\kappa_{\ell n}$ 
is real, i.e. for which the field can propagate through the hole as the $\ell n$ waveguide mode. 
At a fixed illumination wavelength $\lambda_{0}$, this corresponds to cutoff hole diameters given 
by $d_{\ell n}^{\rm c}=x_{\ell n}\lambda_{0}/\pi$. 

The $\ell n$ waveguide mode is one term $J_{\ell}\left(k_{\ell n}\rho\right)e^{i\ell\varphi}e^{i\kappa_{\ell n}z}$ 
of the expansion. For the associated field ${\bf E}^{\rm wg}_{\ell n}({\boldsymbol \rho},z)
=\sum_{i=1,2} \hat{\bf e}_i{\cal E}^{\rm wg}_i(\rho)e^{i\ell\varphi}e^{i\kappa_{\ell n}z}$ with 
$ \hat{\bf e}_{1,2}=(\hat{\boldsymbol \rho},\hat{\boldsymbol \varphi})$, this leads, both for TE and TM polarizations, to a separation between radial and angular 
variables as ${\cal E}^{\rm wg}_i({\boldsymbol \rho})=\hat{\cal E}^{\rm wg}_i(\rho)e^{i\ell\varphi}$. This shows that each $\ell n$ waveguide mode is carrying 
an angular momentum $\ell$: from the fundamental mode ${\rm TE}_{11}$ with $\ell=1$, to the higher 
modes ${\rm TM}_{01}$, ${\rm TE}_{21}$, etc., with respectively $\ell=0$, $\ell=2$, etc. This scaling 
corresponds to OAM cutoff conditions, just as the choice of the cutoff diameter $d_{\ell n}^{\rm c}$ 
selects the $\ell n$ mode. This aspect, only exceptionally discussed in relation with the far field 
transmission properties of cylindrical apertures \cite{Vuong}, becomes key in our experiments and cannot, 
therefore, be discarded.

In this context, the first step is to quantify the excitation efficiency of the waveguide modes inside 
the aperture by the incoming field. This efficiency is determined by an overlap integral 
${\cal O}_{n\ell\ell^{\prime}}=\int_{\rm hole}{\rm d}{\boldsymbol \rho} \ {{\bf E}^{\rm wg}_{\ell n}}^{\star}
({\boldsymbol \rho},z=0^+)\cdot{\bf E}^{\rm exc}_{\ell^{\prime}}({\boldsymbol \rho},z=0^+)$ between the 
in-plane components of the guided ${\bf E}^{\rm wg}_{\ell n}$ field and the excitation ${\bf E}^{\rm exc}_{\ell^{\prime}}$ 
field at the front-side of the membrane, integral performed over the hole. 

The excitation field can either be ${\bf E}^{\rm in}_{\pm}$ or the SP field launched by the grooves with 
${\bf E}^{\rm SP}={\sf C}_{\rm in}(m)\cdot {\bf E}^{\rm in}_{\pm}$. In the former case, because $E_z^{\rm in}=0$, 
only a ${\rm TE}_{11}$ mode is excited, dominating the transmission because of the propagation phase 
$e^{i\kappa_{\ell n}h}$ \cite{JueminPRL2012}. Under normal incidence, only radial or azimuthal polarizations 
can excite higher modes \cite{Leuchs}. In contrast, the launched SP field can excite both TE and TM inside the hole as it is scattered on the edges of the hole. 

At this point, it is essential to refer explicitely to the angular momentum $\ell^{\prime}$ carried by the excitation field which, for ${\bf E}^{\rm in}_{\pm}$, is determined by the spin of light $\ell^{\prime}_s=\pm 1$ and for ${\bf E}^{\rm SP}$ by the spin-orbit coupling in the near field with $\ell^{\prime}_{\rm SP}=m\pm 1$, as shown in Eq.(\ref{matIn}). Through the angular integration, the separation of variables leads to a strict OAM selection rule
\beq
{\cal O}_{n\ell\ell^{\prime}}=\alpha_\ell \left(k_{\ell n}\right)\delta_{\ell,\ell^{\prime}}
\label{overlap}
\eeq
where $\alpha_\ell \left(k_{\ell n}\right)$ equals the integrated value. This selection means that the hole will only sustain the total angular momentum of the in-coupled field if the latter falls within the OAM cutoff conditions determined from the choice of $d_{\ell n}^{\rm c}$. 

The discussion leads to the definition of an effective matrix that connects the front-side ($z=0^+$) and the back-side ($z=h$) of the membrane while simultaneously accounting for OAM transfer from both sides. Such a matrix is defined for each $\ell n$ hole waveguide mode as ${\sf H}_{\ell n}=\sum_i e^{i\kappa_{\ell n}h} {{\cal E}^{\rm wg}_{\ell n}}_i ({\boldsymbol \rho}_{0}) \int\limits_{\rm hole} {\rm d}{\boldsymbol \rho} \ {{\cal E}^{\rm wg}_{\ell n}}^{\star}_i ({\boldsymbol \rho}) \hat{\bf e}_i\otimes\hat{\bf e}_i$ with no cross-terms because of the rotational symmetry of the aperture. 

Because the hole is the optical element in the system that connects both the far and near fields from both sides of the membrane, the whole OAM generation in the far-field transmission process is described through a product-matrix ${\sf T}=\left({\sf C}(m_{\rm out})+{\sf I}_{\rm out}\right)^{\dagger}\cdot {\sf H}_{\ell n}\cdot \left({\sf I}_{\rm in}+{\sf C}(m_{\rm in})\right)$. With ${\sf I}$ proportional to the identity matrix, the ${\sf I}_{\rm out}\cdot {\sf H}_{\ell n}\cdot {\sf I}_{\rm in}$ channel corresponds to the Fano-like path yielding the far-field interferograms. In general, we will assume that when they can be excited, the channels that involve ${\sf C}$ operators are resonantly enhanced over the direct illumination ${\sf I}_{\rm in}$ or direct diffraction ${\sf I}_{\rm out}$ ones.
%%%%%%%%%%%%%%%%%%%%%%%% FIGURE 3 %%%%%%%%%%%%%%%%%%
 \begin{figure}[ht]
 %\vspace{1cm}
 \includegraphics[width=8cm, keepaspectratio] {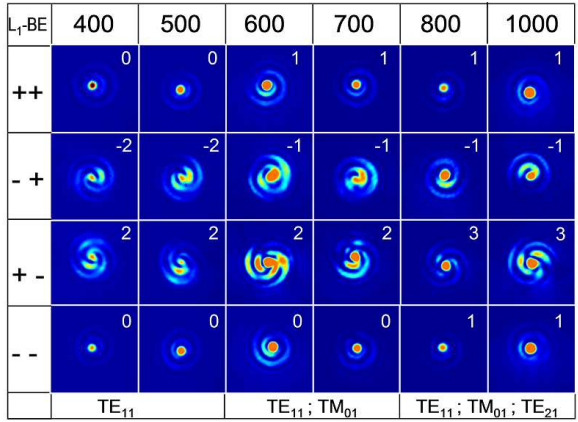}%
 \caption{Intensity (normalized) distributions measured through an $L_1$-BE structure with varying hole diameters $d_h$ from $400$ nm to $1~\mu$m. Smae notations as in Fig.\ref{fig2}. The bottom scale monitors the allowed waveguide modes inside the aperture as a function of $d_h$. }
 \label{fig3}
 \end{figure}
 %%%%%%%%%%%%%%%%%%%%%%%%%%%%%%%%%%%%%%%%%%%%%%%%%%%%%%
Let us analyze precisely the OAM cutoff conditions corresponding to our experimental conditions. 
The fundamental waveguide mode in the central aperture is the ${\rm TE}_{11}$ mode at which the 
hole sustains an OAM index $\ell=1$. At $\lambda_{0}$, we evaluate the cutoff diameter for the 
${\rm TM}_{01}$ at $d_{01}^{\rm c}=580$ nm. For a diameter $d\geq d_{01}^{\rm c}$, the hole can 
thus sustain, in addition to $\ell=1$, another OAM index of $0$. An OAM index of $2$ will be 
allowed above the ${\rm TE}_{21}$ cutoff diameter found at $d_{21}^{\rm c}=740$ nm. A full agreement 
between Table \ref{table} and the experimental results is reached when combining this dimension analysis 
with the selection rules derived above. 

This is seen on the data gathered in Fig.\ref{fig3} for the $L_{1}$-BE structure. The $(i,j)$ 
interferograms monitor the evolution of the OAM generation as a function of cutoff conditions. For 
$d_h=400$ nm, the hole only sustains $\ell=1$, while with the given $m_{\rm in}=-1$ the SP topological 
index will read $\ell_{\rm SP}=m_{\rm in}\pm 1$, equals to $0$ or $-2$. In this case, the OAM transmission is a 
$\left({\sf C}(m_{\rm out})+{\sf I}_{\rm out}\right)^{\dagger}\cdot {\sf H}_{\ell n}\cdot {\sf I}_{\rm in}$ 
process which corresponds to a summation rule in Table \ref{table} with $m_{\rm in}= 0$ and with 
$m_{\rm out}=0$ (the latter being consistent with the BE on the back-side). This analysis solves the 
early paradoxical observations when flipping the BE-$R_1$ structure.

When $d_h$ reaches $600$ nm, the ${\rm TM}_{01}$ mode is allowed, in addition with the ${\rm TE}_{11}$ mode. 
This TM mode can only be excited by the SP field and this time, the selection rule can be fulfilled when 
$\ell = m_{\rm in} + 1= 0$, i.e. when the incident light is $\hat{\boldsymbol \sigma}_+$ polarized. 
In this case, we expect transmitted $\pm 1$ OAM indices in the $(\pm,+)$ subspace. When the incident 
light is $\hat{\boldsymbol \sigma}_-$ polarized, $\ell=m_{\rm in}-1=-2$ is not yet supported by the hole. 
This makes the interferograms $(\pm,-)$ exhibit indices of 2 and 0 which is, again, expected from 
$m_{in}=m_{out}=0$ situation.

However an $\ell=-2$ value is allowed once $d_h\geq 740$ nm. The last two columns of Fig.\ref{fig3} for $d_h=800$ 
nm and $1~\mu$m correspond to holes where ${\rm TE}_{21}$ can be excited. The OAM measurements are fully 
compatible with Table \ref{table} with $m_{\rm in}=1$ and $m_{\rm out}=0$.
%%%%%%%%%%%%%%%%%%%%%%%%% FIGURE 4 %%%%%%%%%%%%%%%%%%
 \begin{figure}[ht]
% \vspace{1cm}
 \includegraphics[width=7.5cm, keepaspectratio] {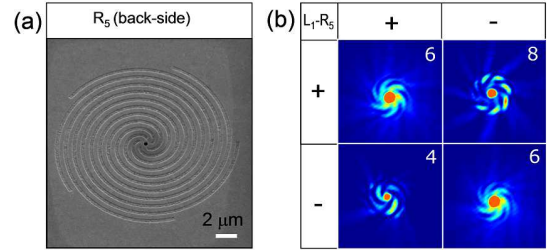}%
 \caption{(a) Scanning electron microscope image of an $R_5$ spiral milled on the back side of a membrane. (b) Intensity (normalized) distributions measured through an $L_1$-$R_5$ structure. Same notations is in Fig.\ref{fig2}.}
 \label{fig4}
 \end{figure}
%%%%%%%%%%%%%%%%%%%%%%%%%%%%%%%%%%%%%%%%%%%%%%%%%%%%%
In short, the OAM cutoff conditions imposed by the central hole take an active part in the far field OAM 
process we have been describing. In this sense, the relation between the hole and the singularity of the 
coupled near field evidences the importance of spin-orbit interaction at the level of single apertures 
\cite{Vuong}. These conditions can be exploited to get a further control over OAM generation as they imply 
specific designs for the front-side structures. To illustrate this most dramatically, we have prepared an 
$(L_{1}-R_{5},d_h=800~{\rm nm})$ structure. As shown in Fig.\ref{fig4} b), the $L_5$ chiral structure consists of 5 
intertwined archimedean spirals. This suspended structure allows generating, through a $TE_{21}$ waveguide mode, OAM in the far field up to $\ell = 8$, again in perfect agreement with the expected OAM summation rules. This value is in strict relation with the chosen structures and is not a limit to our device.

These experiments have revealed important fundamental properties of singular plasmonic interactions 
in the near field. The OAM cutoff properties related to the aperture are essential in understanding the relation between OAM evolutions and reciprocity and they make our devices working like plasmon-based OAM optical diodes \cite{diode}. Also versatility of such devices make them easily integrable in 2D plasmonic systems as optical vortex generators useful for optical communication \cite{CaiScience2012}. We thus believe that the concepts discussed here bear a fundamental importance in plasmonics and nanophotonics as well as provide a basis for novel applications in nanotechnology.  

\indent {\it Acknowledgments - }The authors acknowledge support from the ERC (Grant 227557).


\begin{thebibliography}{99}
%\vspace*{-0.5cm}
\bibitem{HellNatMeth2009} S. W. Hell, Nat. Methods {\bf 6}, 24 (2009).
\bibitem{DemasOptLett2012} J. Demas, M. D. W. Grogan, T. Alkeskjold, and S. Ramachandran, Opt. Lett. {\bf 37}, 3768 (2012). 
\bibitem{Dunlop} M. E. J. Friese, T. A. Nieminen, N. R. Heckenberg, and H. Rubinsztein-Dunlop, Nature \textbf{394}, 348 (1998).
\bibitem{Torres} G. Molina-Terriza, J. P. Torres, and L. Torner, Nat. Phys. \textbf{3}, 305 (2007).
\bibitem{Pendry} J. B. Pendry, Science \textbf{306}, 1353 (2004).
\bibitem{ParkerNatNanotech2007} A. R. Parker and H. E. Townely, Nature Nanotechnol. {\bf 2}, 347 (2007).
\bibitem{GuOptX2011} M. D. Turner, G. E. Schr\"oder-Turk, and M. Gu, Opt. Express \textbf{19}, 10001 (2011).
\bibitem{Zheludev} V. A. Fedotov, P. L. Mladyonov, S. L. Prosvirnin, A. V. Rogacheva, Y. Chen, and N. I. Zheludev, Phys. Rev. Lett. \textbf{97}, 167401 (2006).
\bibitem{BaiPRA2007} B. Bai, Y. Svirko, J. Turunen, and T. Vallius, Phys. Rev. A {\bf 76}, 023811 (2007).
\bibitem{Drezet} A. Drezet, C. Genet, J.-Y. Laluet and T. W. Ebbesen, Opt. Express \textbf{16}, 12559 (2008).
\bibitem{Cohen} Y. Tang and A. E. Cohen, Phys. Rev. Lett. \textbf{104}, 163901 (2010).
\bibitem{HendryNatNano2010} E. Hendry, T. Carpy, J. Johnston, M. Popland, R. V. Mikhaylovskiy, A. J. Lapthorn, S. M. Kelly, L. D. Barron, N. Gadegaard, and M. Kadodwala, Nat. Nanotechnol. {\bf 5}, 783 (2010).
\bibitem{GiessenNanoLett2012} M. Hentschel, M. Sch\"aferling, T. Weiss, N. Liu, and H. Giessen, Nano Lett. {\bf 12}, 2542 (2012).
\bibitem{Babiker} V. E. Lembessis, M. Babiker, and D. L. Andrews, Phys. Rev. A \textbf{79}, 011806 (2009).
\bibitem{Zhan2009} S. Yang, W. Chen, R. L. Nelson, and Q. Zhan, Opt. Lett. {\bf 34}, 3047 (2009).
\bibitem{Halas} S. Zhang, H. Wei, K. Bao, U. H\aa kanson, N. J. Halas,P. Nordlander, and H. Xu, Phys. Rev. Lett. \textbf{107}, 096801 (2011).
\bibitem{Garcia-Vidal} F. R\"uting, A. I. Fern\'andez-Dom\'{\i}nguez, L. Mart\'{\i}n-Moreno, and F. J. Garc\'{\i}a-Vidal, Phys. Rev. B \textbf{86}, 075437 (2012).
\bibitem{ChoOptX2012} S.-W. Cho, J. Park, S.-Y. Lee, H. Kim, and B. Lee, Opt. Express {\bf 20}, 10083 (2012).
\bibitem{Ohno} T. Ohno and S. Miyanishi, Opt. Express \textbf{14}, 6285, (2007).
\bibitem{PRL2008} Y. Gorodetski, A. Niv, V. Kleiner, and E. Hasman, Phys. Rev. Lett. \textbf{101}, 043903 (2008).
\bibitem{Nano2009} Y. Gorodetski, N. Shitrit, I. Bretner, V. Kleiner, and E. Hasman, Nano Lett. \textbf{9}, 3016 (2009).
\bibitem{Zhan} G. Rui, R. L. Nelson, and Q. Zhan, Opt. Lett. \textbf{36}, 4533 (2011).
\bibitem{Beaming_Science} H. J. Lezec, A. Degiron, E. Devaux, R. A. Linke, L. Mart\'{\i}n-Moreno, F. J. Garc\'{\i}a-Vidal, T. W. Ebbesen, Science \textbf{297}, 820 (2002).
\bibitem{DegironOptX2004} A. Degiron and T.W. Ebbesen, Opt. Express {\bf 12}, 3694 (2004).
\bibitem{Fringes} M. Harris, C. A. Hill, and J. M. Vaughan, Opt. Commun. \textbf{106}, 161,(1994).
\bibitem{note1} The direct transmission/diffraction channel can be cancelled by analyzing the transmitted beam in a state cross-polarized with respect to the incident polarization. The interferogram is then replaced by a mere doughnut shaped beam, as shown in Fig.\ref{fig1}c). The determination of $\ell$ on such an image, while still possible, is not as straightforward as from an interferogram. Therefore, we will always project the transmitted beam on a cross-polarized state made slightly elliptical in order to revive the contribution of the direct transmission channel and thus image again the associated intereference.
\bibitem{Allen_book} L. Allen, S. M. Barnett, and M. J. Padgett, \textsl{Optical Angular Momentum} (Institute of Physics, 2003).
\bibitem{LaluetOptX2007} J.-Y. Laluet, E. Devaux,C. Genet, T. W. Ebbesen, J.-C. Weeber,and A. Dereux, Opt. Express {\bf 15}, 3488 (2007).
\bibitem{TeperikOptX2009} T. V. Teperik, A. Archambault, F. Marquier, and J. J. Greffet, Opt. Express {\bf 17}, 17483 (2009).
\bibitem{Manu} E. Lombard, A. Drezet, C. Genet, and T. W. Ebbesen, New J. Phys. \textbf{12}, 023027, (2010).
\bibitem{Brasselet} E. Brasselet, N. Murazawa, H. Misawa, and S. Juodkazis, Phys. Rev. Lett. \textbf{103}, 103903 (2009).
\bibitem{Marucci} L. Marrucci, C. Manzo, and D. Paparo, Phys. Rev. Lett. \textbf{96}, 163905 (2006).
\bibitem{Nikitin} A. Yu. Nikitin, D. Zueco, F. J. Garc\`{i}a-Vidal, and L. Mart\`{i}n-Moreno, Phys. Rev. B \textbf{78}, 165429 (2008).
\bibitem{Catrysse} P. B. Catrysse, H. Shin, and S. Fan, J. Vac. Technol. B \textbf{23}, 2675 (2005).
\bibitem{Vuong} L. T. Vuong, A. J. L. Adam, J. M. Brok, P. C. M. Planken, and H. P. Urbach, Phys. Rev. Lett. \textbf{104}, 083903 (2010). 
\bibitem{JueminPRL2012} J.-M. Yi, A. Cuche, F. de Le\'on-P\'erez, A. Degiron, E. Laux, E. Devaux, C. Genet, J. Alegret, L. Mart\'{\i}n-Moreno, and T. W. Ebbesen, Phys. Rev. Lett. {\bf 109}, 023901 (2012).
\bibitem{Leuchs} J. Kindler, P. Banzer, S. Quabis, U. Peschel, and G. Leuchs, Appl. Phys. B \textbf{89}, 517 (2007).
\bibitem{diode} J. Hwang, M. H. Song, B. Park, S. Nishimura, T. Toyooka, J. W. Wu2, Y. Takanishi, K. Ishikawa, and H. Takezoe, Nat. Mat. \textbf{4}, 383 (2005).
\bibitem{CaiScience2012} X. Cai, J. Wang, M. J. Strain, B. Johnson-Morris, J. Zhu, M. Sorel, J. L. O'Brien, M. G. Thompson, and S. Yu, Science {\bf 338}, 363 (2012).
\end{thebibliography}
\end{document}